\documentclass[12pt,a4paper,draft]{article}
\usepackage{worldsci,xypic}
\LaTeXdiagrams
\let\d\partial
\let\t\widetilde
\let\h\widehat
\def\ad{\dagger}
\def\implies{\;\Longrightarrow\;}
\newtheorem{defn}{Definition}
\newtheorem{thm}{Theorem}

\author{J J C NIMMO\\Department of Mathematics, University of Glasgow,\\
Glasgow G12 8QW, Scotland}
\title{DARBOUX TRANSFORMATIONS FROM REDUCTIONS\\OF THE KP HIERARCHY}
\begin{document}
\pagestyle{empty}
\maketitle
\thispagestyle{empty}
\begin{abstract}
The use of effective Darboux transformations for general classes Lax
pairs is discussed. The general construction of ``binary'' Darboux
transformations
preserving certain properties of the operator, such as self-adjointness,
is given. The classes of Darboux transformations found include the
multicomponent BKP and CKP reductions of the KP hierarchy.
\end{abstract}
\section{Introduction}
Darboux transformations define a mapping between the solutions of a
linear differential equations and a similar equation containing
different coefficients. Since integrable nonlinear evolution equations
frequently arise as the compatibility condition for a pair for such
equations, Darboux transformations may be used to construct families
of exact solutions of the nonlinear equations. Typically these are
multi-soliton solutions. A good introduction to this topic, including
the following example, is given in the monograph by Matveev and
Salle\cite{Matveev&Salle}.

As an example which motivates the work to be presented, consider the Lax
pair\footnote{Here and below $\d=\d/\d x$ and $\d_y=\d/\d y$ and so on.}
\[
L=i\d_y+\d^2+u,\ \ M=\d_t+4\d^3+3\d u+3u\d+3i\d^{-1}(u_y).
\]
for the variant of the Kadomtsev-Petviashvili equation known as KPI
\[
(u_u+6uu_x+u_{xxx})_x)-3u_{yy}=0,
\]
in which $u$ is a real variable. This means that $[L,M]=0$ if and only if
$u$ satisfies KPI.

For all non-zero $\theta$ such that $L(\theta)=M(\theta)=0$, a Darboux
transformation is defined by the operator $G=\theta\d\theta^{-1}$ in the
sense that
$$
L(\psi)=M(\psi)=0\implies\t L(G(\psi))=\t M(G(\psi))=0,
$$
where $\t L$ and $\t M$ are the operators obtained from $L$ and $M$ by
replacing $u$ by $\t
u=u+2(\log\theta)_{xx}$. This result is readily proved by observing that
$$
\t LG=GL\ \ \mbox{and}\ \ \t MG=GM,
$$
i.e.
$$
\t L=GLG^{-1}\ \ \mbox{and}\ \ \t M=GMG^{-1}.
$$
In this way the Darboux transformation manifests itself as a
(differential) gauge transformation.
It also follows that
$$
[L,M]=0\implies[\t L,\t M]=0,
$$
i.e. $\t u$ satisfies KPI whenever $u$ does. Hence the Darboux
transformation induces an auto-B\"acklund transformation.

There is a problem with this transformation however. For almost all $u$,
since $\theta$ is the solution of a complex equation, $\t u$ is not real
and so we do not obtain solutions of KPI. At the root of the problem is
the fact that, while $L$ and $M$ are self-adjoint, $\t L$ and $\t M$ are
not. In order to overcome this problem one may use a ``binary''
transformation (to be defined in the next section) which does preserve
the self-adjointness of $L$ and $M$.

This paper is concerned with the use of binary transformations to
preserve the structure of two classes of operators with matrix
coefficients and arbitrary order.

\section{The structure of the binary transformation}

For an (matrix) operator  $L$, let $S=\{\theta,\
\text{non-singular}:L(\theta)=0\}$
(and define $\t S$, $S^\ad$ for operators $\t L$, $L^\ad$ etc.). A
(formally invertible) gauge transformation $G_\theta$, for $\theta\in
S$, defines a mapping
$$
G_\theta\colon S\to \t S,\ \ \text{where}\ \ \t L=G_\theta
LG_\theta^{-1}.
$$

Consider also the (formal) adjoint operator $G_\theta^\ad$. (Taking the
formal adjoint is, as usual, the linear operation defined by
$$
(a\d^i)^\ad=(-1)^i\d^ia^\ad,
$$
for a matrix $a$, where $a^\ad$ denotes the Hermitian conjugate of $a$.)
Since $\t L^\ad=G_\theta^{\ad^{-1}}L^\ad G_\theta^\ad$, we have
$$
G_\theta^\ad\colon \t S^\ad\to S^\ad.
$$

By determining the kernel of $G_\theta^\ad$ we obtain some nontrivial
solution in $\t S^\ad$. Typically, we may identify this subset in terms
of $\theta$ and denote a member as $i(\theta)$. For example, in the
classical case when $G_\theta=\theta\d\theta^{-1}$ and
$G_\theta^\ad=-\theta^{\ad^{-1}}\d\theta^\ad$, we find that
$$
G_\theta^\ad(\rho)=0\iff\rho=\theta^{\ad^{-1}}c,
$$
where $c$ is independent of $x$.

We represent this situation in the diagram below.
$$
\begin{diagram}
S\rto^{G_\theta}\xdotted[1,1]+<-2mm,2mm>|<{\rotate\tip}|>\tip
\save\go+<0mm,-5mm>\drop{\scriptstyle\theta}\restore&\t S\\
S^\ad&\lto_{G_\theta^\ad}\t S^\ad\save\go+<0mm,-5mm>
\drop{\scriptstyle i(\theta)}\restore
\end{diagram}.
$$

To describe the general form of the binary transformation we consider
operators $L$, $\t L$ and $\h L$ and the corresponding sets of
non-singular solutions matrices $S$, $\t S$ and $\h S$. Let $\theta\in
S$ and $\h\theta\in \h S$ be such that $G_\theta\colon S\to\t S$ and
$G_{\hat\theta}\colon \h S\to\t S$. Then we get the mapping
$$
G_{\hat\theta}^{-1}G_\theta^{\phantom{\hat\theta}}\colon S\to\h S.
$$

The difficulty with this definition of a transformation is that to
define it we need one of the solutions we are trying to determine,
namely $\h\theta$! To overcome this, we use the fact that there
corresponds to $\h\theta\in\h S$ a solution $i(\h\theta)\in\t S^\ad$ and
then use
the mapping $G_\theta^{\ad^{-1}}\colon S^\ad\to\t S^\ad$ to obtain
$\h\theta=i^{-1}(G_\theta^{\ad^{-1}}(\phi))$ for any $\phi\in S^\ad$.
This is shown in the diagram below.
$$
\spreaddiagramrows{-1pc}
\begin{diagram}
S\rto^{G_\theta}&\t S&\lto_{G_{\hat\theta}}\h
S\xdotted[1,-1]!<5mm,0mm>|<{\rotate\tip}|>\tip\\
S^\ad&\lto_{G^\ad_\theta}\t S^\ad&\\
\save\go+<0mm,2mm>\Drop{\phi}\xto[0,1]+<-3mm,2mm>|<\stop\restore&
\save\go+<1mm,2mm>\Drop{i(\h\theta)}\restore&
\end{diagram}
$$

In this way we obtain the definition of a general binary transformation.
\begin{defn} \upshape
Consider an operator $L$ and gauge operator $G_\theta$, where
$\theta\in S$, such that $G_\theta^\ad(i(\theta))=0$. For each
$\phi\in S^\ad$, define
$$
G_{\theta,\phi}=G_{\hat\theta}^{-1}G_\theta^{\phantom{\hat\theta}},
$$
where $\h\theta=i^{-1}(G_\theta^{\ad^{-1}}(\phi))$. Then
$$
G_{\theta,\phi}\colon S\to \h S,
$$
where $\h L=G_{\theta,\phi}^{\phantom{-1}}LG_{\theta,\phi}^{-1}$, is
called a \emph{binary transformation}.
\end{defn}
In the next section we will consider two concrete examples of such a
binary transformation.

Now suppose that the operator $L$ has a constraint of the form
$$
L^\ad R^\ad=RL,
$$
where $R$ is in some formally invertible (matrix differential)
operator\footnote{
It is tempting to look for a constraint of the form $L^\ad S=RL$ but
this in fact corresponds to two constraints since on taking adjoints
$L^\ad R^\ad=S^\ad L$.}.
We wish to find binary transformations that preserve this constraint.
That is---using the notation of the above definition---we want $\h L$ to
satisfy the constraint whenever $L$ does.

Examples for the choice of $R$ include
\begin{itemize}
\item $R=I$. $L$ is self-adjoint. An example of the application of this
is quoted in the introduction.
\item $R=iI$. $L$ is skew-adjoint. This corresponds to the CKP reduction
of the KP hierarchy\cite{Date et al 1} and the reduction of the
Kuperschmidt ``$k=0$'' non-standard
hierarchy\cite{Kuperschmidt,Konopelchenko&Oevel}.
\item $R=\d$. This corresponds to the BKP\cite{Date et al 2} or the
Kuperschmidt ``$k=1$'' reduction.
\end{itemize}
The binary transformation we will discuss in the next section will preserve
generalizations of these three reductions.

Let the gauge transformation $G$, such that $\h L=GLG^{-1}$,
preserve the constraint $L^\ad=RLR^{\ad^{-1}}$. Then
$$
\h L^\ad-R\h LR^{\ad^{-1}}=0
$$
which means that
$$
G^{\ad^{-1}}L^\ad G^\ad-RGLG^{-1}R^{\ad^{-1}}=
G^{\ad^{-1}}RLR^{\ad^{-1}}G^\ad-RGLG^{-1}R^{\ad^{-1}}=0.
$$
This leads to the single condition $RG=G^{\ad^{-1}}R$.

Note that the relation between $L$ and its adjoint imposes a
relationship between the solution sets $S$ and $S^\ad$. In particular,
for each $\theta\in S$, $R^\ad(\theta)\in S^\ad$. Hence,
in the case of a binary transformation
$G=G_{\theta,\phi}$, we may make the choice $\phi=R^\ad(\theta)$.

\section{Darboux transformations for general operators}

In this section we describe two classes of Darboux transformation for
general classes of matrix differential operators of arbitrary order. The
first is originally due to Matveev\cite{Matveev} and has also been
considered recently by Oevel\cite{Oevel}. We will present a very simple
proof of this result. The second was found by Oevel
\& Rogers\cite{Oevel&Rogers} in the case of scalar operators in the
context of Sato theory. We will derive a more general version here.

In both cases, the results are remarkably general. There is however a
serious drawback. There is, in this general case, absolutely no
guarantee that the transformed operator we have the same
``form'' as the original and so only in special cases does one get a
transformation that induces an auto-B\"acklund transformation.

First, consider
$$
L=\d_t+\sum_{i=0}^nu_i\d^i\ \ \mbox{and}\ \ \t L=\d_t+\sum_{i=0}^n\t
u_i\d^i,
$$
where $u_i$ and $\t u_i$ are $N\times N$ (not necessarily constant)
matrices. Let the operator $G$ be such that
$$
\t L=GLG^{-1}=L+[G,L]G^{-1}.
$$
Hence $G$ must satisfy
$$
[G,L]G^{-1}=\sum_{i=0}^n (\t u_i-u_i)\d^i.
$$
Taking $G=\theta\d\theta^{-1}$, where $\theta$ is a non-singular
$N\times N$ matrix, and hence $G^{-1}=\theta\d^{-1}\theta^{-1}$, we get
$$
[G,L]G^{-1}=[G,L]\theta\d^{-1}\theta^{-1}=
\sum_{i=0}^n a_i\d^{i-1}\theta^{-1},
$$
for some matrices $a_i$. For $i=1,\ldots,n$, $a_i=\t u_{i-1}-u_{i-1}$
and in order that $G$ define a Darboux transformation we must have
$$
a_0=0.
$$
This condition gives $[G,L](\theta)=0$ i.e. $G(L(\theta))=0$ since
$G(\theta)=0$. Hence we only need require that $L(\theta)=\theta C$, for
some $x$-independent matrix $C$. Note that if $L(\theta)=0$ then for
$\theta'=\theta\exp(\d_t^{-1}(C))$, $L(\theta')=\theta'C$. Also,
$G_{\theta'}=G_\theta$ and so we may suppose, without loss of
generality, that $C=0$\footnote{Note
that if $L$ is an ordinary differential operator, then taking $C\ne0$ is a
genuine generalization. For example, this is exploited in the classical
``discrete eigenvalue adding'' Darboux transformation for the
time-independent Schr\"odinger operator.}. Thus we find that $\theta\in
S$.

The second case we consider is
$$
L=\d_t+\sum_{i=1}^nu_i\d^i\ \ \text{and}\ \ \t L=\d_t+\sum_{i=1}^n\t
u_i\d^i,
$$
where $u_i$ and $\t u_i$ are again $N\times N$ matrices.
Note that the multiplicative term in $L$ and $\t L$ is omitted.

As in the first case, a gauge operator $G$ must satisfy
$$
[G,L]G^{-1}=\sum_{i=1}^n\t u_i-u_i.
$$
There are now two simple choices. First, let $G=G_\theta^{(1)}=\theta^{-1}$, an
(invertible) $N\times N$ matrix. Then
$$
[G,L]G^{-1}=\sum_{i=0}^n a_i\d^i,
$$
and so $a_0=[G,L](\theta)=G(L(\theta))=0$, i.e. $\theta\in S$.

Second, let $G=G^{(2)}_\rho=\rho_x^{-1}\d$, where $\rho_x$ is an invertible
$N\times N$ matrix. Now
$$
[G,L]G^{-1}=[G,L]\d^{-1}\rho_x=\sum_{i=1}^{n+1} a_i\d^{i-1}\rho_x,
$$
and we must have $a_1=0$, i.e. $[G,L](\d^{-1}(\rho_x))=G(L(\rho))=0$.
Thus $L(\rho)=C$, an $x$-independent matrix. Again, we may take $C=0$
without loss of generality and so $\rho\in S$.

As in the scalar case\cite{Oevel&Rogers}, it
is the composition of the two gauge transformations which is of most
interest, and we take
$G_\theta=G^{(2)}_{G^{(1)}_\theta(1)}G^{(1)}_\theta=(\theta^{-1})_x^{-1}
\d\theta^{-1}$.

\section{Binary transformations and reductions}

To determine the binary transformations $G_{\theta,\phi}$ corresponding to
the two Darboux
transformations found above we must determine two additional things: the
mapping $i\colon \hat S\to\t S^\ad$ and then the element $\h\theta\in\h
S$ in terms of $\theta$ and $\phi$.

First consider $L=\d_t+\sum_{i=0}^nu_i\d^i$, $G_\theta=\theta\d\theta^{-1}$.
Here the condition
$G_\theta^\ad(i(\theta))=-\theta^{\ad^{-1}}\d(\theta^\ad i(\theta))=0$
is satisfied by the choice $i(\theta)=\theta^{\ad^{-1}}$. Further,
\begin{eqnarray*}
\h\theta&=&\left(G_\theta^{\ad^{-1}}(\phi)\right)^{\ad^{-1}}\\
        &=&-\left(\theta^{\ad^{-1}}\d^{-1}(\theta^\ad\phi)\right)^{\ad^{-1}}\\
        &=&-\theta\Omega^{-1},
\end{eqnarray*}
where $\Omega=\d^{-1}(\phi^\ad\theta)$. It may be shown that for all
operators $L=\sum_{i=0}^nu_i\d^i$, $\Omega$ is
exact in the sense that $d\Omega=\phi^\ad\theta
dx+A(u_1,\ldots,u_n,\theta,\phi)dt$\cite{Oevel}.

In this case the binary transformation is
\begin{eqnarray*}
G_{\theta,\phi}=G_{\hat\theta}^{-1}G_\theta^{\phantom{-1}}
&=&\theta\Omega^{-1}\d^{-1}\Omega\d\theta^{-1}\\
&=&\theta\Omega^{-1}\d^{-1}(\d\Omega-\Omega_x)\theta^{-1}\\
&=&1-\theta\Omega^{-1}\d^{-1}\phi^\ad.
\end{eqnarray*}
For discussion of the reduction we will also need
$$
G^{\ad^{-1}}=1-\phi\Omega^{\ad^{-1}}\d^{-1}\theta^\ad.
$$

Now suppose that $L$ satisfies the constraint $L^\ad R^\ad=RL$ where
$R=A$, a (not necessarily constant) matrix. Then we may choose
$\phi=R^\ad(\theta)=A^\ad\theta$. The condition
$RG_{\theta,\phi}=G_{\theta,\phi}^{\ad^{-1}}R$ now gives
\begin{eqnarray*}
A-A\theta\Omega^{-1}\d^{-1}\phi^\ad=A-\phi\Omega^{\ad^{-1}}\d^{-1}\theta^\ad A
&\iff&A\theta\Omega^{-1}\d^{-1}\theta^\ad
A=A^\ad\theta\Omega^{\ad^{-1}}\d^{-1}\theta^\ad A\\
&\iff&A^\ad=\pm A.
\end{eqnarray*}

This establishes the following theorem.
\begin{thm}\label{0}
Let the matrix operator $L=\sum_{i=0}^n u_i\d^i$ satisfy the constraint
$$
L^\ad A=AL,
$$
where $A$ is an Hermitian or skew-Hermitian matrix. Then the binary
transformation
$$
G=1-\theta\Omega^{-1}\d^{-1}\theta^\ad A
$$
where $\Omega=\d^{-1}(\theta^\ad A\theta)$, preserves the above
constraint, i.e. $\h L=GLG^{-1}$ satisfies $\h L^\ad A=A\h L$.
\end{thm}

For the second case, $L=\sum_{i=1}^n u_i\d^i$,
$G_\theta=(\theta^{-1})_x^{-1}\d\theta^{-1}$ and hence
$i(\theta)=(\theta^{\ad^{-1}})_x$.

To determine the binary transformation it is notationally convenient to
write an element of $S^\ad$ as $\phi_x$ rather than $\phi$ as we did
above. Also, it is necessary to introduce two integrals
$$
\Omega=\d^{-1}(\phi^\ad\theta_x)\ \ \mbox{and}\ \
\Omega'=\d^{-1}(\phi_x^\ad\theta),
$$
where
$$
\Omega+\Omega'=\phi^\ad\theta.
$$
Now
\begin{eqnarray*}
i(\h\theta)=(\h\theta^{\ad^{-1}})_x
        &=&G_\theta^{\ad^{-1}}(\phi_x)\\
        &=&-(\theta^{\ad^{-1}})_x\d^{-1}(\theta^\ad\phi_x)
\end{eqnarray*}
and so
$$
(\h\theta^{-1})_x=-\d^{-1}(\phi^\ad_x\theta)(\theta^{-1})_x=
-\Omega'(\theta^{-1})_x.
$$
Integrating by parts and taking inverses, we get
\begin{eqnarray*}
\h\theta&=&\left(-\Omega'\theta^{-1}+\d^{-1}(\Omega'_x\theta^{-1})\right)^{-1}\\
        &=&\left(\phi^\ad-\Omega'\theta^{-1}\right)^{-1}\\
        &=&\theta\Omega^{-1}.
\end{eqnarray*}

We may now obtain
\begin{eqnarray*}
G_{\theta,\phi_x}
&=&\h\theta\d^{-1}(\h\theta^{-1})_x(\theta^{-1})_x^{-1}\d\theta^{-1}\\
&=&-\theta\Omega^{-1}\d^{-1}\Omega'\d\theta^{-1}\\
&=&1-\theta\Omega^{-1}\d^{-1}\phi^\ad\d,
\end{eqnarray*}
and in a similar way
$$
G_{\theta,\phi_x}^{\ad^{-1}}=1-\d\phi\Omega'{}^{\ad^{-1}}\d^{-1}\theta^\ad.
$$

Suppose that $L$ satisfies the constraint $L^\ad R^\ad=RL$ where
$R=A\d$, $A$ a matrix, and choose
$\phi_x=R^\ad(\theta)=-(A^\ad\theta)_x$, i.e. $\phi=-A^\ad\theta$. The
condition $RG_{\theta,\phi_x}=G_{\theta,\phi_x}^{\ad^{-1}}R$ is
$$
A\d-A\d\theta\Omega^{-1}\d^{-1}\theta^\ad A\d=A\d-
                 \d A^\ad\theta\Omega'{}^{\ad^{-1}}\d^{-1}\theta^\ad A\d
\iff A^\ad=\pm A\ \mbox{and}\ A_x=0.
$$
With these conditions on $A$, $\Omega=\pm\Omega'{}^\ad$.

This establishes a second theorem.
\begin{thm}\label{1}
Let the matrix operator $L=\sum_{i=1}^n u_i\d^i$ satisfy the constraint
$$
L^\ad A\d+A\d L=0,
$$
where $A$ is an $x$-independent Hermitian or skew-Hermitian matrix.
Then the binary transformation
$$
G=1-\theta\Omega^{-1}\d^{-1}\theta^\ad A\d
$$
where\footnote{For a better notation we have replaced $\Omega$ with
$-\Omega$ in the statement of the theorem} $\Omega=\d^{-1}(\theta^\ad
A\theta_x)$, preserves the above
constraint, i.e. $\h L=GLG^{-1}$ satisfies $\h L^\ad A\d+A\d\h L=0$.
\end{thm}

\section{Examples}

\subsection{Davey-Stewartson I}
This system has Lax pair
$$
L=\d_y+\alpha\d+Q,\ \ M=i\d_t+\alpha\d^2+\frac12(Q\d+\d Q+\alpha Q_y)+D,
$$
where $Q=\left(\begin{array}{cc}0&u\\\epsilon\bar u&0\end{array}\right)$
and $D=\left(\begin{array}{cc}U&0\\0&V\end{array}\right)$ is real.

Let $A=\left(\begin{array}{cc}1&0\\0&-\epsilon\end{array}\right)$. Then
$$
L^\ad(iA)^\ad=(iA)L,\ \ M^\ad A^\ad=AL.
$$
Hence we may use Theorem~\ref{0} (with $A=I$) to obtain a binary
transformation.
This transformation has been used to obtain a wide class of solutions
including dromions\cite{Nimmo1}.

\subsection{Sawada-Kotera equation}
The equation is a reduction of the BKP equation and so has Lax pair
admitting the BKP reduction:
$$
L=(\d^2+3u)\d,\ \ M=\d_t+(9\d^5-15\d
u\d+30(\d^2u+u\d^2)+15u^2)\d,
$$
where
$$
L^\ad\d+\d L=M^\ad\d+\d M=0.
$$
We may use Theorem~\ref{1} (with $A=I$) to obtain the binary
transformation. Note that this is given by
$G=\theta^{-1}\d^{-1}\theta^2\d\theta^{-1}$ and coincides the with the
well-known ``Darboux'' transformation\cite{AiyerFuchsteiner&Oevel,Nimmo2}.

\subsection{Modified Novikov-Veselov equation}
This system\cite{Konopelchenko(book),Nimmo3,Oevel&Schief}
belongs to the two component BKP hierarchy and has a Lax pair
$$
L=\d_y+S\d,\ \ M=\d_t+(S\d^2+T\d+\d T+U)\d,
$$
where $S=\left(\begin{array}{cc}\cos u&\sin u\\\sin u&-\cos
u\end{array}\right)$, and $T$ and $U$ are give skew-symmetric and
symmetric real matrices respectively. Again
$$
L^\ad\d+\d L=M^\ad\d+\d M=0
$$
and Theorem~\ref{1} gives the binary transformation.

\section{Conclusions}
We have discussed the general construction of binary transformations
from Darboux transformations. In particular we have carried this out for
two classes of operators. More importantly, we have shown that the well
studied reductions of these classes (the multi-component BKP and CKP
reductions) are among those that the binary transformations preserve.

Iteration of the binary transformation is, of course, possible and leads
to closed-form expressions for solutions---of the linear problems and
for the integrable systems that are their compatibility conditions---in
terms of (multi-component) Grammian determinants. In the case of the
reduction described in Theorem~\ref{1}, one may see how these Grammians
are transformed into Pfaffians by the reduction process. These features
will be discussed in more details elsewhere\cite{Nimmo4}.

\section{Acknowledgements}
I wish to thank Ralph Willox and Walter Oevel for useful discussions
relating to this work and the Royal Society of London for financial
support enabling my attendance at this workshop.

\end{document}